\documentclass[%
 aip,
rsi,%
 amsmath,amssymb,
 reprint,%
]{revtex4-1}
\usepackage{graphicx}
\usepackage{dcolumn}
\usepackage{bm}
\usepackage{subcaption}
\usepackage{caption}
\usepackage[labelfont=bf,textfont=it,labelsep=endash]{subcaption} 
\usepackage{stackengine}
\usepackage{placeins}
\usepackage{color}
\makeatletter
\newcommand\makesimplesubcaption{\caption@@@make{\caption@fnum{sub\@captype}}}
\makeatother

\begin{document}

\preprint{AIP/123-QED}

\title[Herapath et al]{Impact of pump wavelength on terahertz emission of a cavity-enhanced spintronic trilayer}

\author{R.I. Herapath}
 \altaffiliation{rh522@exeter.ac.uk}
 
\author{S.M. Hornett}%
\affiliation{ 
University of Exeter, Physics building, Stocker Road, Exeter, UK, EX4 4QL
}%

\author{T.S. Seifert}
\affiliation{%
Fritz Haber Institute of the Max Planck Society, 14195 Berlin, Germany
}%
\affiliation{%
Department of Physics, Freie Universit\"at Berlin, 14195 Berlin, Germany
}%

\author{G. Jakob}
\affiliation{%
Institute of Physics, Johannes Gutenberg University Mainz, 55099 Mainz, Germany
}%
\affiliation{%
Graduate School of Excellence Materials Science in Mainz, Staudinger Weg 9, 55128 Mainz, Germany
}%

\author{M. Kl\"aui}
\affiliation{%
Institute of Physics, Johannes Gutenberg University Mainz, 55099 Mainz, Germany
}%
\affiliation{%
Graduate School of Excellence Materials Science in Mainz, Staudinger Weg 9, 55128 Mainz, Germany
}%

\author{J. Bertolotti}%

\affiliation{ 
University of Exeter, Physics building, Stocker Road, Exeter, UK, EX4 4QL
}%
\author{T. Kampfrath}

\affiliation{%
Fritz Haber Institute of the Max Planck Society, 14195 Berlin, Germany
}%
\affiliation{%
Department of Physics, Freie Universit\"at Berlin, 14195 Berlin, Germany
}%

\author{E. Hendry}%
 
\affiliation{ 
University of Exeter, Physics building, Stocker Road, Exeter, UK, EX4 4QL  
}%

\date{\today}

\begin{abstract}
We systematically study the pump-wavelength dependence of terahertz pulse generation in thin-film spintronic THz emitters composed of a ferromagnetic Fe layer between adjacent nonmagnetic W and Pt layers. We find that the efficiency of THz generation is essentially flat for excitation by 150~fs pulses with center wavelengths ranging from 900 to 1500~nm, demonstrating that the spin current does not depend strongly on the pump photon energy. We show that the inclusion of dielectric overlayers of TiO\textsubscript{2} and SiO\textsubscript{2}, designed for a particular excitation wavelength, can enhance the terahertz emission by a factor of of up to two in field.

\end{abstract}

\pacs{Valid PACS appear here}
\maketitle
\raggedbottom
\begin{figure}[t]
\includegraphics{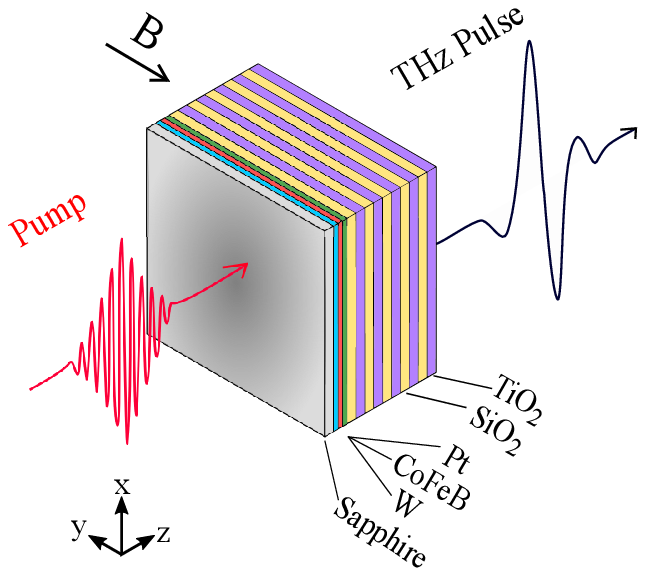}

\caption[]{\label{fig:schematic}Schematic of a spintronic trilayer with added dielectric cavity, grown on 0.5~mm of sapphire (Al\textsubscript{2}O\textsubscript{3}). The near-infrared pump pulse, incident through the substrate, is partially absorbed in the metallic layers, launching a spin current from the ferromagnetic (FM) layer into the nonmagnetic (NM) layers. The inverse spin Hall effect converts this ultrashort out-of-plane spin current into an in-plane charge current resulting in the emission of THz radiation into the optical far-field. A weak in plane magnetic field (B) determines the magnetization direction, and the linear polarization of the emitted THz field.}
\end{figure}

Terahertz (THz) radiation is non-ionizing and, therefore, safe for many applications, ranging from micro- and macroscopic imaging and spectroscopy to wireless communication\cite{Pawar2013,Federici2005,Sizov2010,Dhillon2017}. However, the spectroscopically interesting THz frequency band near 1~THz is not easily accessible. Electronic sources such as oscillators can only provide high (milliwatt) output levels up to a few 100~GHz \cite{Zhang2010}, while optical sources such as quantum cascade lasers are typically limited to frequencies \textgreater  2 THz at room temperature \cite{Lewis2014a,Dhillon2017}. To fill this ``gap", considerable effort has been garnered towards sources capable of frequency mixing and optical rectification \cite{Burford2017}, typically driven by femtosecond lasers. 

To date, most THz emitting materials have been found to be insulators or semiconductors \cite{Lee2009}. Recently, THz emitters based on magnetic, metallic thin films have been demonstrated which emit THz radiation under illumination by femtosecond pulses\cite{Seifert2016a,Huisman2017,Yang2016Heterostructure,Papaioannou,Feng2018,Torosyan2017,Wu2017}. We here focus on trilayer thin-film emitters formed from a ferromagnetic (FM) layer between two non-ferromagnetic (NM) layers. A two-step process is thought to generate THz radiation:\cite{Kampfreth2013TerahertzHeterostructures} Upon excitation by the femtosecond pump pulse, an ultrashort out-of-plane spin current polarized along the FM magnetization is injected from the FM into the NM layers. Thereafter, the inverse spin Hall effect converts the laser-induced spin current into a transverse in-plane charge current within the NM layer which leads to the emission of a terahertz pulse into the optical far-field \cite{Kampfreth2013TerahertzHeterostructures,Battiato2010,Saitoh2006}. 

One of the most efficient films of this type\cite{Seifert2016a} comprises W, CoFeB and Pt layers. Importantly, Pt and W feature a spin Hall angle of opposite sign, resulting in a constructive superposition of the two charge currents in these layers. The result is an ultrabroadband THz emitter, capable of delivering  pulses spanning 0.1 to 30~THz.\cite{Seifert2016a} With an active region only a few nanometers thick in total, these emitters can generate as much THz radiation as a phase-matched electrooptic crystal of millimeter thickness \cite{Seifert2016a}. Such highly efficient, but thin, THz emitters show much promise, particularly for near-field measurement or applications that benefit from the absence of phase matching. 

Most studies of these THz emitters have been carried out using Ti:sapphire laser sources with wavelengths around 800~nm. However, many thin metal films show a rather wavelength-independent absorptance in the visible and near infrared, such that the THz-generation efficiency may naively be expected to be largely independent of the pump wavelength \cite{Seifert2016a}. Such wavelength-independent emission would be a great advantage of these types of emitters, allowing users free choice in excitation laser source.

In this paper, we investigate the pump-wavelength dependence of THz emission of W$|$CoFeB$|$Pt trilayer emitters using a continuously tuneable femtosecond laser source. We find that the efficiency of THz generation is surprisingly flat for excitation by 150~fs pulses with central wavelengths ranging from 900 to 1500~nm. This observation reveals that the photon energy has little effect on the number of electrons contributing to transport of spin polarization and that the key parameter is the total amount of energy deposited by the pump pulse. We demonstrate that the inclusion of dielectric overlayers (TiO\textsubscript{2} and SiO\textsubscript{2}), forming a cavity with the substrate, can enhance emission by up to a factor of four in intensity in the frequency window from 0 to 2~THz.

\begin{figure*} 
\begin{subfigure}{0.42\textwidth}
\phantomcaption
\stackinset{l}{2.4in}{b}{1.4in}{\makesimplesubcaption{}}
{\includegraphics[width=\linewidth]{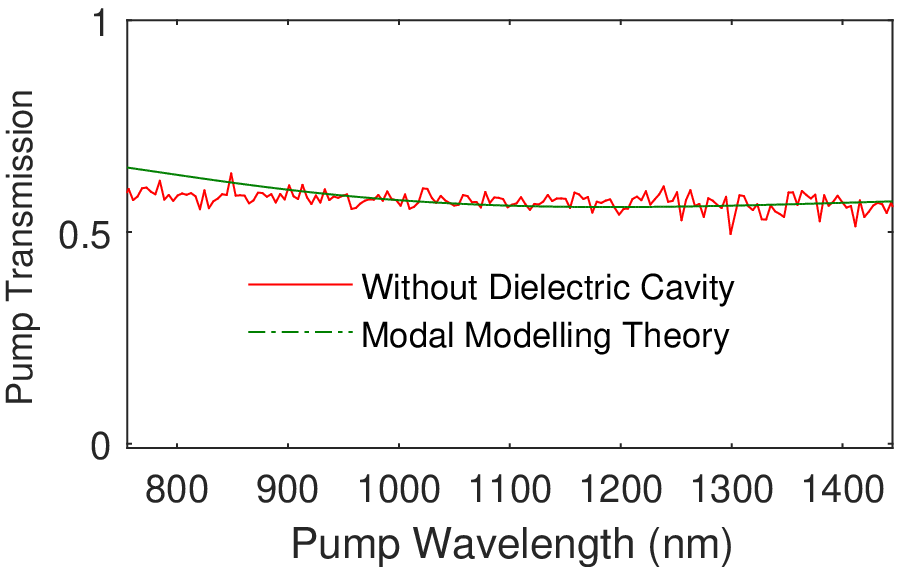}}
 \label{fig:mononrm}
\end{subfigure}
\begin{subfigure}{0.43\textwidth}
\phantomcaption
\stackinset{l}{2.43in}{b}{1.4in}{\makesimplesubcaption{}}
{\includegraphics[width=\linewidth]{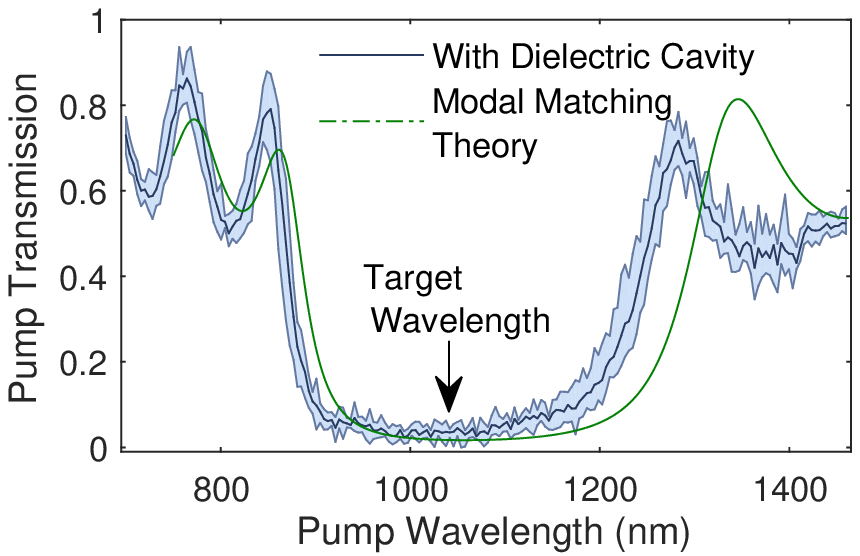}}
\label{fig:monodiele}
\end{subfigure}\\[-3ex]
\smallskip
\begin{subfigure}{0.42\textwidth}
\phantomcaption
\stackinset{l}{2.4in}{b}{1.4in}{\makesimplesubcaption{}}
{\includegraphics[width=\linewidth]{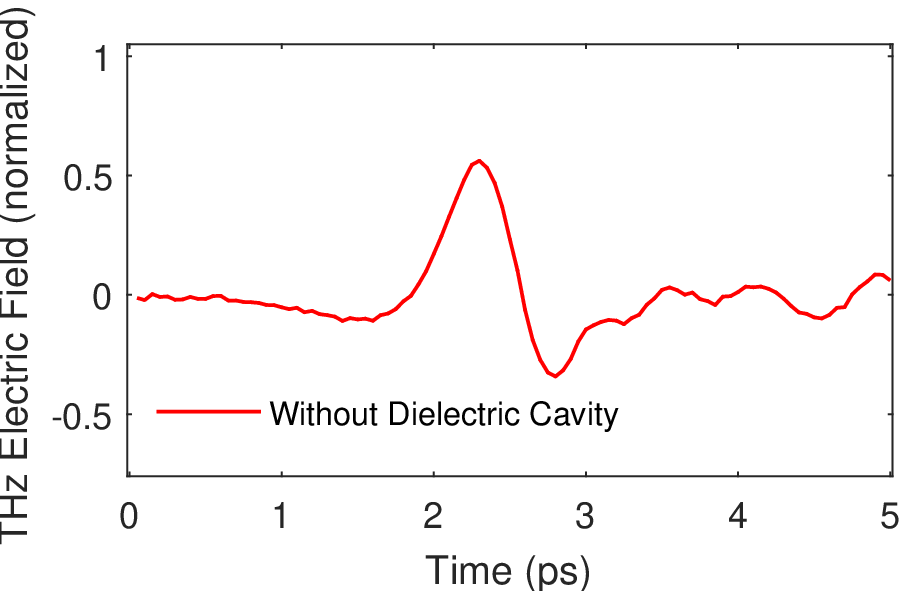}}
\label{fig:tracenrm}
\end{subfigure}
\begin{subfigure}{0.42\textwidth}
\centering
\phantomcaption
\stackinset{l}{2.4in}{b}{1.4in}{\makesimplesubcaption{}}
{\includegraphics[width=\linewidth]{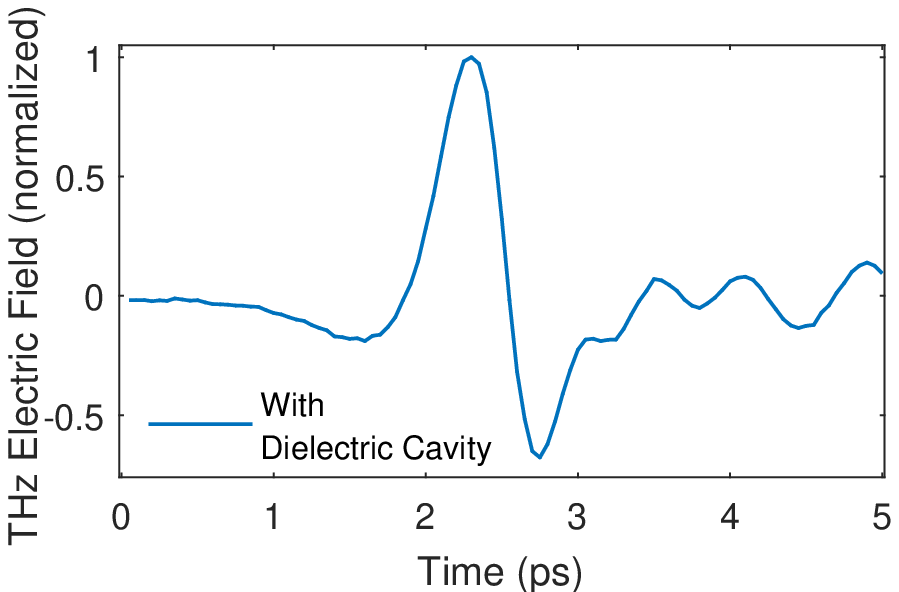}}
\label{fig:tracediele}
\end{subfigure}\\[-3ex]
\hspace{-5.10mm}
\smallskip
\begin{subfigure}{0.41\textwidth}
\centering
\phantomcaption
\stackinset{l}{2.4in}{b}{1.6in}{\makesimplesubcaption{}}
{\includegraphics[width=\linewidth]{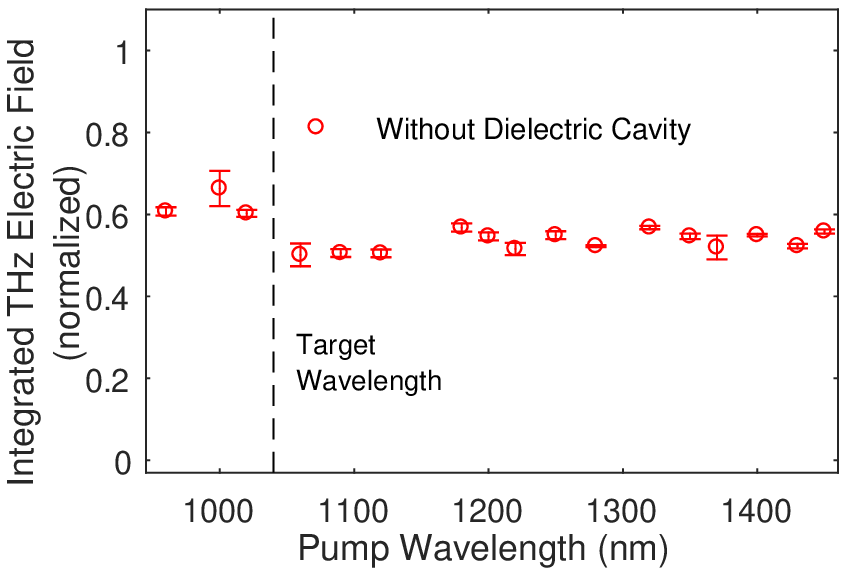}}
\label{fig:wavenrm}
\end{subfigure}
\begin{subfigure}{0.42\textwidth}
\centering
\phantomcaption
\stackinset{l}{2.5in}{b}{1.65in}{\makesimplesubcaption{}}
{\includegraphics[width=\linewidth]{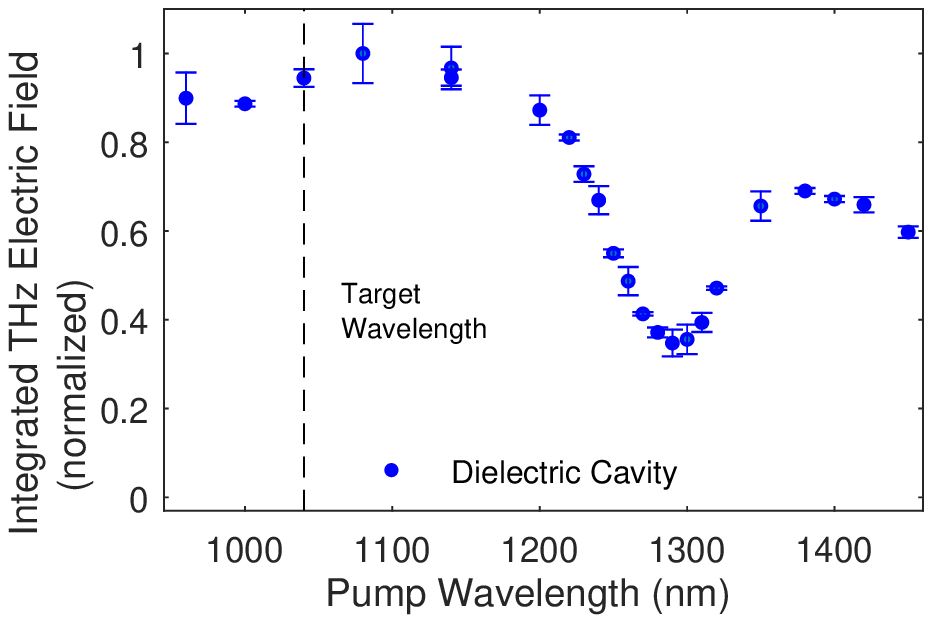}}
\label{fig:wavediele}
\hspace{6.1mm}
\end{subfigure}\\[-3ex]
\caption{(a, b) Measured transmittance of samples without (left) and with (right) with TiO\textsubscript{2} and SiO\textsubscript{2} dielectric overlayers. The blue-shaded region in panel~(b) represents variation among the six cavity films due to variation in the fabrication process. Whilst the transmittance of the trilayer alone is relatively flat with incident wavelength, the trilayer with the dielectric overlayers exhibits a bandstop region between 900 and 1200~nm. Green lines are calulated using modal matching (see supplementary information). (c, d)~Typical THz waveforms from the bare trilayer sample (left) and the trilayer with overlayers (right)  pumped by 1040~$\mu$m pulses. (e, f)~The THz electric field, integrated over the spectrum, emitted without (left) and with (right) dielectric overlayers as function of the pump wavelength while the energy, focus diameter and duration of the pump pulses were kept constant.  Error bars represent variation in emission between nominally identical samples. Note that the highest value of THz field has been normalized to one to ease comparison. } 
\end{figure*}

A schematic of the samples is shown in Fig. \ref{fig:schematic}. The emitters have each 10~mm by 7~mm in area, with layers of W, CoFeB and Pt deposited on a sapphire substrate by Ar sputtering using a Singulus Rotaris\textsuperscript{\copyright} deposition system.\cite{Seifert2016a} The layers of W, CoFeB and Pt have a nominal thickness of 2~nm, 1.8~nm and 2~nm, respectively, giving a total thickness of 5.8~nm for the metallic layers. On six samples, an optical cavity made from alternating dielectric overlayers of TiO\textsubscript{2} (thickness 113~nm, near-infrared index of refraction 2.265)\cite{Artimis} and SiO\textsubscript{2} (185~nm, 1.455), with a total overlayer thickness of 1.495~$\mu$m, was deposited by plasma-assisted electron beam evaporation using a Leybold 1104 coating platform. Note that the overlayer thicknesses were optimized numerically to give a flat response at the target wavelength. The near-infrared transmission spectrum, normalized by the optical transmission of the sapphire substrate, was recorded for each sample using a monochromator (Oriel) and Quartz Tungsten-Halogen lamp (Thorlabs) (see Figs. \ref{fig:mononrm} and \ref{fig:monodiele}).

THz emission was investigated using wavelength-tunable near-infrared pump pulses (duration $\sim 150$~fs, spot diameter $\sim 4$~mm, fluence $\sim 0.1$~J~m\textsuperscript{-2}) incident normal to the sample surface. They were generated by an optical parametric amplifier (Light Conversion TOPAS) driven by an 800~nm Ti:Sapphire amplified laser ($\sim 100$~fs, repetition rate of 1~kHz). When varying the pump wavelength, we paid particular attention to keeping the other pump-pulse parameters, in particular energy, duration and focus diameter, constant. The resulting THz pulse, emitted in the forward direction, is detected through electrooptic sampling with 800~nm pulses from our amplified laser system using a 1~mm thick, (110)-oriented ZnTe crystal. For all samples, a neodymium magnet attached to the emitter holder approximately 7~mm from the edge of the film gives an in-plane magnetic field of $\sim 13$~mT, as depicted in Fig. \ref{fig:schematic}. This determines the linear polarization of the emitted THz field which is always oriented perpendicularly to the magnetic field.

Figure \ref{fig:tracenrm} shows a typical electrooptic signal waveform of a THz pulse emitted from the bare spintronic trilayer after excitation by a pump pulse with a center wavelength of 1040~nm. The signal amplitude grows approximately linearly for incident pump fluences at least up to 0.4~J~m\textsuperscript{-2} (see Supplementary Fig. \ref{fig:powdep}).

In Fig. \ref{fig:wavenrm}, we plot the integrated THz field emitted by a W$|$CoFeB$|$Pt trilayer sample for various pump wavelengths in the range from 900 to 1500~nm. Remarkably, over this range, the efficiency of THz generation is wavelength-independent. Importantly, since they originate from the same source, the focus diameter, energy and duration of the pump pulses are fairly independent of the wavelength in the range 1000 to 1300~nm. For example, the pulse duration varies by less than 12~fs across this entire range (see Supplementary Fig. \ref{fig:autocorr} and Supplementary Section \ref{sec:length}). An unchanged THz signal amplitude has also been reported\cite{Papaioannou} for the two significantly shorter wavelengths of 800~nm and 400~nm. Our results and those of Ref. \cite{Papaioannou} indicate that the spin current arises from the hot electrons induced by the pump pulse. Details of the involved optical transitions are insignificant. Therefore, the key parameter is the amount of energy that is deposited by the pump pulse in the electronic system.

Since the THz emission from the spintronic emitters is largely independent of pump wavelength, one can in principle enhance the emission for any particular wavelength by designing a suitable adjacent cavity. A straightforward implementation are dielectric overlayers similar to a Bragg mirror, forming a broadband dielectric cavity with the substrate. By placing the lower index material, in this case SiO\textsubscript{2}, next to the trilayer, an intensity maximum is formed within the active THz generation layers. Doing so, for a particular target pump wavelength, pump absorption in the metallic trilayer will be maximized, and so will the emitted THz radiation. For convenience, we choose a target pump wavelength of 1040~nm aimed at common fibre lasers, specified by the thickness and type of the dielectric layers. To achieve a stop-band with less than 5\% transmission, a total of 10 dielectric layers are required. For these coated samples, the total thickness (neglecting the substrate) is 1.495~$\mu$m, corresponding to an optical thickness for THz wavelengths of approximately 9~$\mu$m, which is still very subwavelength for THz radiation.  

To model the behavior of the emitter with dielectric cavity, a standard modal matching approach \cite{Tomas1995} is used which can also treat a source within the active region (see Supplementary Section \ref{sec:modal}). In Fig. \ref{fig:monodiele}, we plot the calculated near-infrared transmission spectrum for the samples with dielectric overlayers (green dot-dash line). We expect a transmission minimum  (Fig. \ref{fig:monodiele}) and an absorptance maximum, (see Supplementary Section \ref{sec:modal}) at the target wavelength of 1040~nm, with a bandwidth of around 300~nm. 

In Fig. \ref{fig:monodiele}, the blue-shaded region represents the distribution of transmittance spectra measured for the six samples with overlayers. While the transmittance (Fig. \ref{fig:monodiele}) and absorptance (Supplementary Fig. \ref{fig:transabs}) of the sample without the dielectric overlayers are fairly flat over the measured range, all samples with dielectric overlayers show clear band-stop behavior. The minimum transmittance in the measurement, $\sim 3$\%, is consistent with our modeling. The slight shift in wavelength of the band-stop region between the modeling and experiment likely arises from variation in the refractive indices and/or thicknesses of the dielectric layers. For example, the optical thickness needs to change by only 3.8\% with respect to specified values to account for this discrepancy. The band-stop in the region of the target wavelength designates a region where we expect to see enhancement of THz emission.

In Fig. \ref{fig:wavediele}, we can observe the enhancement in THz emission generated by the dielectric overlayers. For an identical pump fluence, in the band-stop window (900 to 1200~nm), we find a THz field that is a factor of $\sim 2$ (intensity of a factor $\sim 4$) larger compared to the trilayer without overlayers. This is slightly larger than the 70\% enhancement in field recently reported for a cavity formed from spintronic layers separated by dielectric spacers. \cite{Feng2018} We also observe reduction in THz emission near the near-infrared transmittance maxima at 1300~nm. In these bandpass regions, the dielectric overlayers act to reduce the near-infrared intensity in the active magnetic layer. For the band-stop region, the factor of two enhancement in THz field emission is observed for the entire fluence range investigated here (Supplementary Fig. \ref{fig:powdep}). 

\begin{figure}[t!]
\includegraphics{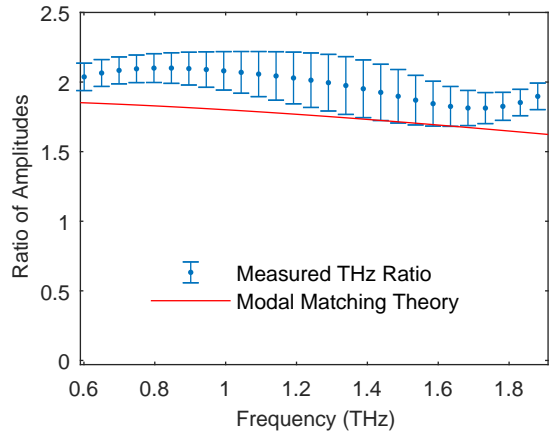}
\caption{\label{fig:THzDep} Measured THz emission enhancement for different spectral components, plotting the THz field amplitude emitted from a spintronic trilayer with the dielectric cavity divided by field emitted without cavity, when excited by a pump pulse of 1040~nm. The solid line is a calculation from modal matching theory.}
\end{figure}

In Fig. \ref{fig:THzDep}, we plot the spectral dependence of this observed THz emission enhancement for excitation by a 1040 nm pump pulse (blue dots). The highest enhancement is observed for the lowest THz frequencies. Above 1~THz, we begin to see a slight drop in the field enhancement factor (limited to 2 THz by the bandwidth of our detector). Also shown is the prediction from modeling (red line): this is calculated from the modal modeling method using the THz field emitted with and without the dielectric cavity, multiplied by the predicted pump intensity in the active magnetic layer. Our modeling predicts a similar behavior, but with an enhancement factor which decreases more quickly with frequency, an effect which arises due to the frequency dependent absorption of TiO\textsubscript{2}. In our model, we also observe high sensitivity to the precise index of the Pt layer, a material which shows high variation dependent on morphology. \cite{Kovalenko2001ThicknessSemiconductors} It is important to note that alternative dielectrics to TiO\textsubscript{2} may well exhibit lower absorption losses, and could increase the operational bandwidth of cavity enhanced emitters. 

To conclude, we systematically study the pump-wavelength dependence of THz emission of spintronic W$|$CoFeB$|$Pt trilayers. We find that the efficiency of THz generation is essentially flat for excitation by 150~fs pulses with central wavelengths ranging from 900 to 1500~nm, indicating that the spin current is largely independent of the pump-photon energy. We demonstrate that the inclusion of dielectric overlayers of TiO\textsubscript{2} and SiO\textsubscript{2}, designed for a particular excitation wavelength, can enhance emission by up to a factor of two in field amplitude. The four-fold enhancement in emitted THz intensity could be further improved using cavities with higher quality factors, matched to the bandwidth of the pump pulse.

The authors like to acknowledge support via the EPSRC Centre for Doctoral Training in Metamaterials (Grant No. EP/L015331/1). EH acknowledges support from EPSRC fellowship (EP/K041215/1). TK, TSS, MK and GJ acknowledge the German Research Foundation for funding through the collaborative research centers SFB TRR 227 Ultrafast spin dynamics (project B02) and SFB TRR 173 Spin+X as well as the Graduate School of Excellence Materials Science in Mainz (MAINZ, GSC 266). TK also acknowledges funding through the ERC H2020 CoG project TERAMAG/Grant No. 681917.

\nocite{*}
\bibliography{MyColl}

\clearpage
\section{Supplementary Material}
\newcommand{\beginsupplement}{%
        \setcounter{table}{0}
        \renewcommand{\thetable}{S\arabic{table}}%
        \setcounter{figure}{0}
        \renewcommand{\thefigure}{S\arabic{figure}}%
        \setcounter{section}{0}
     }
\beginsupplement
\raggedbottom
\begin{figure}
\includegraphics{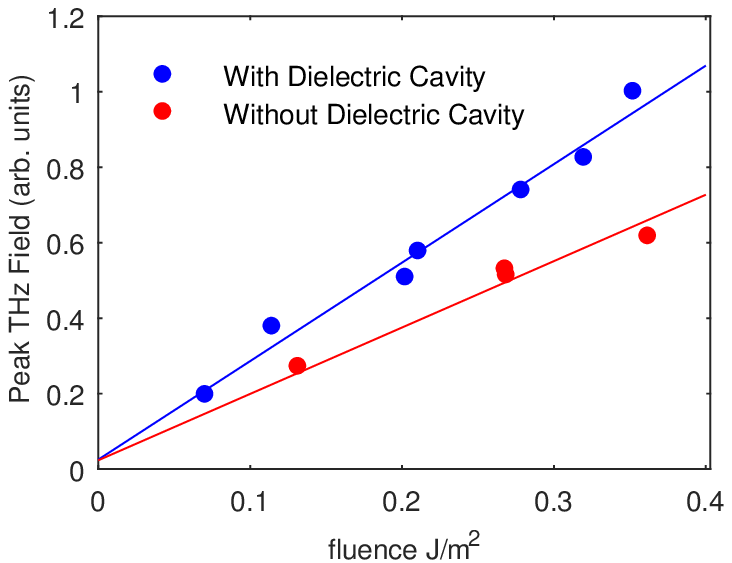}
\caption{\label{fig:powdep} THz emission as function of pump fluence, measured with a pump wavelength of 1000~nm. Both films exhibit linear dependences over this range of fluence, with the dielectric cavity enhancing the THz field emitted. Solid lines are linear fits.}
\end{figure}

\begin{figure}[t!]
\includegraphics[width = 0.45\textwidth]{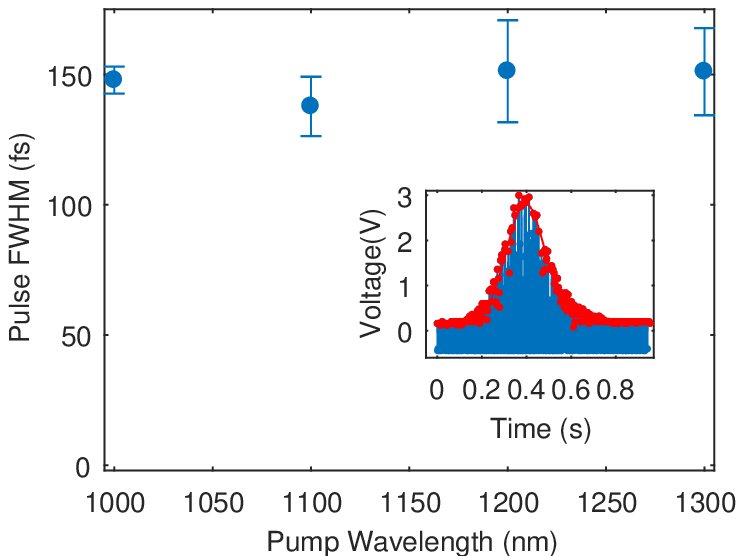}
\caption{\label{fig:autocorr} Variation of pulse lengths as a function of pump wavelength. The full width half maximum (FWHM) in autocorrelation is approximately 150~fs. The inset shows a sample pulse taken at 1200~nm: blue region is the raw data, iteratively thresholded, while the red line is a Gaussian fit.}
\end{figure}

\begin{figure}
\begin{subfigure}{0.4\textwidth}
\phantomcaption
\stackinset{l}{2.3in}{b}{1.7in}{\makesimplesubcaption{}}
{\includegraphics[width=\linewidth]{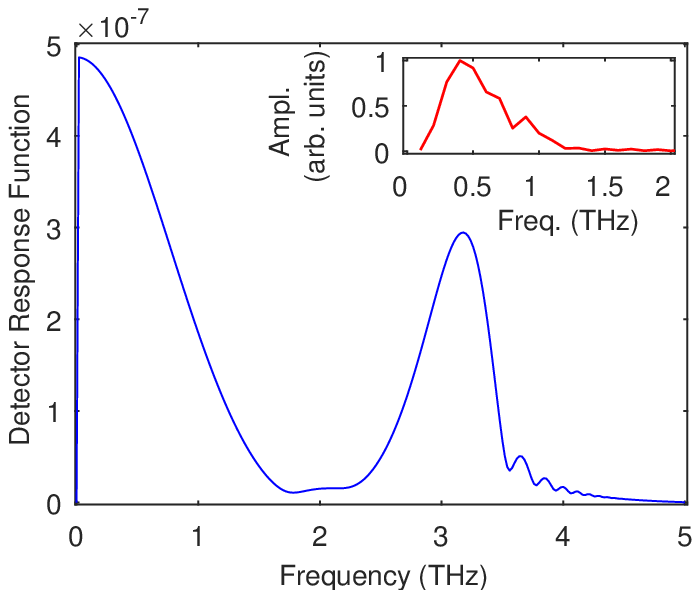}}
\label{fig:detresp1}
\end{subfigure}

\smallskip
\begin{subfigure}{0.4\textwidth}
\phantomcaption
\stackinset{l}{2.3in}{b}{1.7in}{\makesimplesubcaption{}}
{\includegraphics[width=\linewidth]{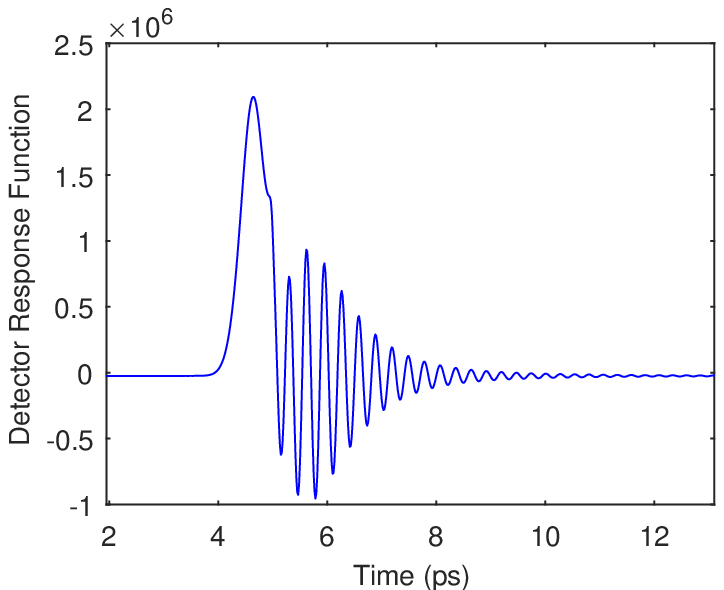}}
\label{fig:detresp2}
\end{subfigure}\\[-4ex]
\caption{The figures above show the modeled detector response function for electrooptic sampling in (a)~the time domain and (b)~the frequency domain. The inset in (a)~ shows the spectrum of the spintronic emitter with dielectric cavity.}
\label{fig:detresp}
\end{figure}

\begin{figure*}
\begin{subfigure}{0.42\textwidth}
\phantomcaption
\stackinset{l}{2.4in}{b}{1.4in}{\makesimplesubcaption{}}
{\includegraphics[width=\linewidth]{Monographnormv5.eps}}
\label{fig:mononrmsup}
\end{subfigure}
\begin{subfigure}{0.434\textwidth}
\phantomcaption
\stackinset{l}{2.45in}{b}{1.4in}{\makesimplesubcaption{}}
{\includegraphics[width=\linewidth]{Monographv4.eps}}
\label{fig:monodielesup}
\end{subfigure}\\[-3ex]
\smallskip
\begin{subfigure}{0.42\textwidth}
\phantomcaption
\stackinset{l}{2.4in}{b}{1.4in}{\makesimplesubcaption{}}
{\includegraphics[width=\linewidth]{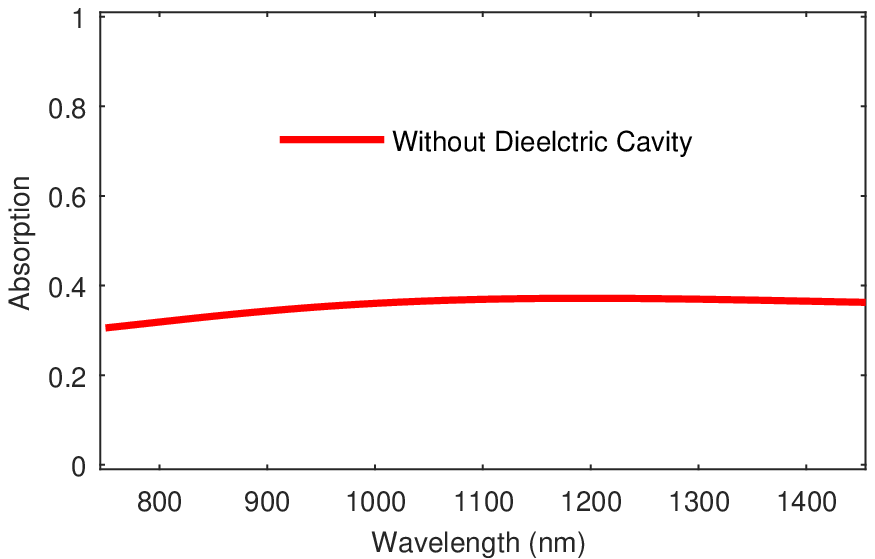}}
\label{fig:absnrm}
\end{subfigure}
\begin{subfigure}{0.42\textwidth}
\phantomcaption
\stackinset{l}{2.4in}{b}{1.4in}{\makesimplesubcaption{}}
{\includegraphics[width=\linewidth]{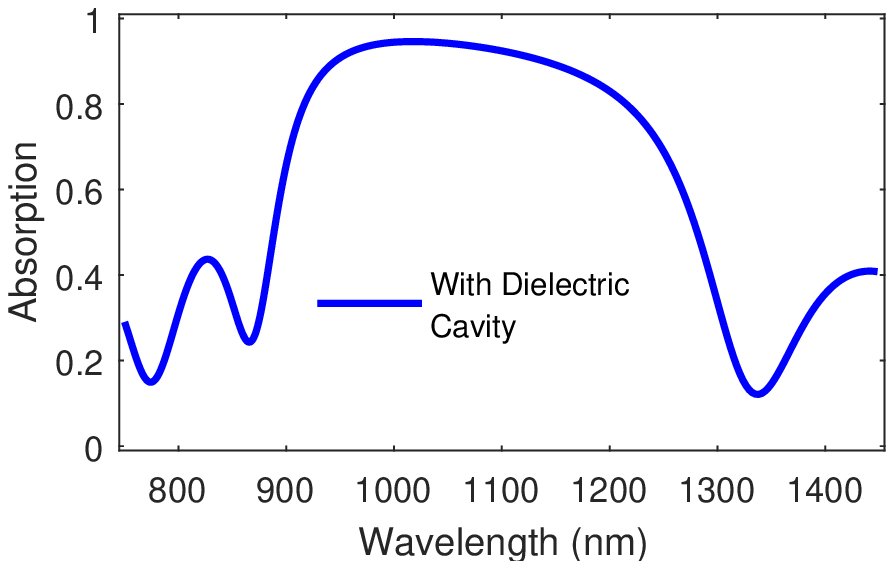}}
\label{fig:absdiele}
\end{subfigure}\\[-4ex]
\caption{(a, b) The transmission of the spintronic emitter without (left) and with (right) the dielectric cavity shows the effectiveness of the dielectric cavity in preventing the pump wavelengths from transmitting through the emitter. (c, d) The absorption of the emitter without (left) and with (right) the dielectric cavity, obtained using the modal matching, shows an increase over the bandstop range. This increased absorption is responsible for the increase in THz radiation when excited in this range.}
\label{fig:transabs}
\end{figure*}

\subsection{\label{sec:powdep}Power dependence of THz emission}
We examine the power dependence of THz emission from the spintronic trilayer with and without the dielectric cavity attached. As seen in Fig. \ref{fig:powdep}, the THz signal amplitude grows linearly with respect to the pump fluence for both samples (with and without cavity) over the entire range investigated here. Non-reversible saturation was observed for pump fluences $\gg 1$~J~cm\textsuperscript{-2}, possibly due to heating of the sample above the Curie temperature of the FM layer and ablation of the metal films .\cite{Seifert2016a,Yang2016Heterostructure}

\subsection{\label{sec:length}Pump-pulse duration vs wavelength} 
We determine the length of the pulses generated by our optical parametric amplifier as a function of wavelength, 1000 to 1300~nm, using an autocorrelator (FR-103MC from Femtochrome). Autocorrelation curves are fitted with a Gaussian, yielding the pulse duration (full width at half intensity maximum). In Fig. \ref{fig:autocorr}, we see that the variation in pulse lengths across our measureable wavelength range is negligible and smaller than the noise in the measurement. Error bars represent variations in measurement, most likely resulting from environmental factors, including humidity and laser fluctuations.

\subsection{\label{sec:detresp}Detector response function}
In Fig.\ref{fig:detresp}, we plot the detector response function $h(t)$ of our THz electrooptic detection system (1~mm thick, (110)-ZnTe crystal and 100~fs, 800~ nm sampling pulses) calculated using Kampfrath et al.\cite{Kampfrath2007SamplingCrystals} It connects the THz electric field incident $E_{\textnormal{inc}}(t)$ onto the electrooptic crystal with the electrooptic signal $S(t)$ by the convolution $S=h*E_{\textnormal{inc}}$. Since $h(t)$ is much wider than the pump pulses ($\sim 150$~fs), we expect that the small variations of the pump-pulse duration present in Fig. \ref{fig:autocorr} are invisible to the electrooptic detection.

\subsection{\label{sec:modal}Modal matching calculations}
The layers are considered homogeneous in the $x$-$y$ plane, where the radiation propagates along the $z$-axis. We define the in-plane component of the electric field in the semi-infinite vacuum region as a sum of forward and backward propagating plane waves of frequency {$\omega$} given by

\begin{equation}
\label{eqn:appa}
A_i \textnormal{e}^{\textnormal{i}n_i\omega z/c} + B_i \textnormal{e}^{-\textnormal{i}n_i\omega z/c}, 
\end{equation}
where $A_i$ and $B_i$ are the amplitudes of the forward- and backward-propagating fields in the $i$-th layer, with refractive index $n_i$, and $c$ being the vacuum speed of light. Note that the forward- and backward-propagating amplitudes in the reflection and transmission regions are zero, respectively. Using Maxwell's equations, one can then find a similar expression for the magnetic field:
\begin{equation}
\label{eqn:appa2}
\frac{\omega}{c}A_in_i \textnormal{e}^{\textnormal{i}{n_i\omega z}/{c}} - \frac{\omega}{c}B_in_i \textnormal{e}^{-\textnormal{i}{n_i\omega z}/{c}}.
\end{equation}
Then, by applying field continuity boundary condition at the interfaces between all the layers, one obtains a set of simultaneous equations that can be solved for the sets of unknowns $A_i$ and $B_i$, and, therefore, for the field amplitudes in each of the regions. We use this model to calculate 1)~the near-infrared transmission of our multilayer stack (by including a source electric field of unit amplitude in the incident region), and 2)~the normalized THz emission of the multilayer stack (by including a unitary source field in the magnetic layer of the stack), both normalized by the field within the sapphire substrate. Literature values for the  metallic\cite{Seifert2016a} and dielectric\cite{Grischkowsky1990,Dang2014ElectricalDeposition} THz refractive indices are used. The optical frequency parameters for TiO\textsubscript{2} and SiO\textsubscript{2} were provided by our commercial fabrication partners \cite{Artimis}.

In Fig. \ref{fig:transabs}, we plot the calculation absorption spectra for our samples without (bold red line) and with (bold blue line) dielectric cavity. Without cavity, the absorption is fairly wavelength-independent. With the cavity, absorption is increased by around a factor of two in the band-stop region of the spectrum. In this region, we also see a commensurate increase in the the THz field emitted from samples with the cavity, as seen in Fig. \ref{fig:wavediele}. 

\end{document}